\documentclass[conference]{IEEEtran}
\usepackage{graphicx}
\usepackage{amsmath}
\usepackage{amssymb}
\usepackage{fancyhdr}
\newlength{\zeroheight}
\begin{document}


\title{
\Large \bf Synchrophasor monitoring of single line outages via area angle and susceptance}
\author{
\IEEEauthorblockN{ Atena Darvishi\hspace{2cm}Ian Dobson}
\IEEEauthorblockA{Electrical and Computer Engineering Department\\
Iowa State University,
Ames IA USA \\
darvishi@iastate.edu, dobson@iastate.edu\\[-5pt]}}
\fancyhead[c]{\textnormal{\small North American Power Symposium (NAPS), September 2014, Pullman WA  USA}}
\renewcommand{\headrulewidth}{ 0.0pt}
\fancyfoot[C]{\fontfamily{ptm}\selectfont\fontsize{10}{10}\null}



%
\maketitle

\thispagestyle{fancy}
\begin{abstract}  
Synchrophasor monitoring of angles around the area has been used to track the area stress caused by the single outage. We applied the idea of the area angle which is a combination of
 synchrophasor measurements of voltage angles around the border of the area to measure the severity of the single line outages inside the area. 
Both idealized and practical examples are given to show that 
the variation of the area angle for single line outages can be approximately related to changes in the overall susceptance of the area 
and the line outage severity. 
\end{abstract}


\section{Introduction}
\label{intro}

The  angle across an area of a power system  is a weighted combination of synchrophasor measurements of voltage phasor angles around the border of the area \cite{DobsonvoltPS12,DobsonIREP10}.
 The weights are  calculated from a DC load flow model of the area in such a way that 
 the area angle satisfies circuit laws.
 Area angles were first developed for the special case of areas called cutset areas that extend all the way across the power system
  \cite{DobsonHICSS10,DobsonPESGM10,LopezPESGM12}.
  We previously showed how area angle responded to single line outages inside the area in some Japanese test cases \cite{DarvishiNAPS13}.
 The increase in area angle largely reflected outage severity and ways to choose the area were discussed.
 
 \looseness=-1
 Area angles are easy to calculate from synchrophasor measurements, and their general value is in 
 giving a fast and meaningful bulk measure related to stress in a specific area of the power system.
 Area angle monitoring would complement slower monitoring via state estimation.
 The approximate relation of changes in area angle to outage severity suggests that it could be 
 easier to set alarm thresholds using area angles.
Another  measure of stress, the voltage angle between two synchrophasor locations, responds to events throughout the power system, 
and is not easy to relate to a particular area.
This paper seeks to quantify outage severity with bulk area monitoring; to identify the line outage in the area see \cite{SehwailNAPS12,SehwailPS13,TatePS08}.

The area angle is measured across an area from one ``side" of the area such as the north to the other side of the area such as the south.
The area susceptance across the area can also be defined, and, according to Ohm's law in a DC power flow context, the equivalent power 
flow through the area is the product of the area susceptance and the area angle.
The power flow through the area is often approximately constant, so it is 
intuitively plausible that when a line outages, the area susceptance decreases and the area angle increases.
In this paper, we explain and examine this approximate relationship between area angle and area susceptance in detail,
including testing on two areas of the WECC. We choose these areas of the transmission system between major generation and major load
   to try to describe with the area angle the stress resulting from the 
   transfer of power through the area from generation to load. There are some arts to choose a good area to be meaningful with respect to power flow direction.
The testing on the WECC areas also shows how changes in area angle can usually distinguish the single line outage severity.
This paper is limited to single line outages that, for simplicity, do not island the system.\footnote{Islanding line outages require assumptions about generator redispatch}
 
 \section{Area angle and area susceptance formulas and relations}
 
\subsection{Formulas for voltage angle and power through the area}
\label{formulas}
We summarize from \cite{DobsonvoltPS12} formulas related to the area angle and power entering the area.
    We consider a connected area R of the power system with border buses $M$ and interior buses $N$.
     The susceptance matrix from the base case DC power flow is written as $B$, with subscripts indicating submatrices or elements of $B$.
      The following notation is used for column vectors of voltage angles and powers:
      
	\vspace{5pt}
\noindent
  \begin{tabular}{ @{\hspace{1.5cm}}ll @{}}
  $\theta_n$&voltage angles at interior buses $N$\\
 $P_n$&power injected at interior buses $N$\\
 $\theta_m$&voltage angles at border buses $M$\\
  $P_m$& power injected at border buses $M$\\
$P_{m}^{\rm into}$&power entering R at border buses $M$\\&\qquad along tie lines\\
   \end{tabular}
   \vspace{5pt}

%
%
 The vector of powers $P_m^{\rm R}$ entering the border buses of R is the sum of the 
 power $P_m$ injected directly at the border buses and the power $P_m^{\rm into}$ flowing into the area along the tie lines:
\begin{align}
P_m^{\rm R}=P_m+P_m^{\rm into}.
\label{ImR}
\end{align}
The susceptance matrix of the area $R$, considered as an isolated area without its tie lines, is $B_{mm}^{\rm R}$.
Retaining the border buses $M$ and applying to $R$ a standard Ward or Kron reduction to eliminate the interior buses $N$, we get
\begin{align}
P_m^ {\rm Rred}&=P^{\rm R}_m- B_{{m}n}B_{nn}^{-1}P_{n},
\label{pmred}\\
B_{mm}^ {\rm Rred}&=
B_{mm}^{\rm R}-B_{mn}B_{nn}^{-1}
B_{nm}.
\label{bmmred}
\end{align}
We indicate the partition of the border buses into two sets $M_a$ and $M_b$ by specifying the row vector $\sigma_a$,
whose $i$th component is one if bus $i$ is in $M_a$, and is zero otherwise.

Now we can define our main quantities. 
An equivalent power \cite{DobsonvoltPS12} that flows 
from $M_a$ to $M_b$ through R is
\begin{align}
P_{\rm area}=\sigma_a  P_m^ {\rm Rred}.
\label{parea}
\end{align}
The susceptance of the area $b_{\rm area}$ is
\begin{align}
b_{\rm area}=\sigma_a  B_{mm}^ {\rm Rred} \sigma_a ^T.
\label{bab}
\end{align}
The area angle $\theta_{\rm area}$ is the scalar quantity
\begin{align}
\theta_{\rm area}&=\frac{\sigma_a  B_{mm}^ {\rm Rred} \theta_m}{b_{\rm area}}\notag\\
&=w  \theta_m
=w[1]  \theta_{m}[1]+w[2]  \theta_{m}[2]+...+w[k]  \theta_{m}[k]
\label{thetanew}
\end{align}
where $w$ is a row vector of weights $w=(w[1],w[2],...,w[k])$ that depend only on the 
area topology and the susceptances of lines in the area. $k$ is the number of border buses.
To monitor the area angle  with (\ref{thetanew}), we use the synchrophasor measurements of $\theta_m$ at the 
border buses and recent base case susceptances and topology of a DC load flow\footnote{Such DC load load flows are generally available \cite{TatePS08}.
}  of the area R to calculate the weights $w$.
If an outage of line $i$ occurs, then the synchrophasor measurements at the border buses change to $\theta_m^{(i)}$ but we 
continue to use the weights computed {\sl before} the outage to compute the area angle as
\begin{align}
\theta_{\rm area}^{(i)}=\frac{\sigma_a B_{mm}^ {\rm Rred} \theta_m^{(i)}}{b_{\rm area}}=w\theta_m^{(i)}.
\label{pmuareaanglei}
\end{align}

\subsection{Approximate inverse relation between area angle and area susceptance}

We informally explain why the monitored area angle $\theta_{\rm area}^{(i)}$ varies approximately inversely to the area susceptance $b_{\rm area}^{(i)}$.
It turns out\footnote{This approximation will be established with more rigor in a future paper.} that the monitored area angle $\theta_{\rm area}^{(i)}$ of (\ref{pmuareaanglei}) is close to 
the following area angle (note the square brackets in the superscript $[i]$):
\begin{align}
\theta_{\rm area}^{[i]}=\frac{\sigma_a B_{mm}^ {{\rm Rred}(i)} \theta_m^{(i)}}{b_{\rm area}^{(i)}}=\frac{\sigma_a B_{mm}^ {{\rm Rred}(i)}\theta_m^{(i)}}{\sigma_a B_{mm}^ {{\rm Rred}(i)}\sigma_a^T }
,\label{pmuareaangleiii}
\end{align}
which is the area angle that would be computed after the outage of line $i$ if the 
outage of  line $i$ were accounted for in the weights. (The difference between (\ref{pmuareaangleiii}) and (\ref{pmuareaanglei})
is that the susceptance matrix  $B_{mm}^ {{\rm Rred}(i)}$ that accounts for the outage of line $i$ replaces $B_{mm}^ {\rm Rred}$ in 
both the numerator and denominator of (\ref{pmuareaangleiii}).)
The results in section \ref{results} show numerical evidence that 
$\theta_{\rm area}^{(i)}$ and $\theta_{\rm area}^{[i]}$ are close; that is,
\begin{align}
\theta_{\rm area}^{(i)}\approx \theta_{\rm area}^{[i]}.
\end{align}
It is the case \cite{DobsonvoltPS12} that Ohm's law applies to area angles so that 
\begin{align}
P_{\rm area}=b_{\rm area} \theta_{\rm area}.
\label{ohm}
\end{align}
In particular, when line $i$ outages, we have 
 \begin{align}
 P^{(i)}_{\rm area}=b^{(i)}_{\rm area} \theta^{[i]}_{\rm area},
 \label{ohmi}
 \end{align} and, from (\ref{ImR}), (\ref{pmred}), and (\ref{parea}), we have 
  \begin{align}
 P^{(i)}_{\rm area}=\sigma_a (P_m+P_m^{{\rm into}(i)}- B_{mn}^{(i)}{(B^{(i)}_{nn}})^{-1}P_{n}).
 \label{pareaiex}
 \end{align}
 Since line $i$ is assumed to be a non-islanding outage, and there are assumed to be no losses in the DC load flow approximation, there is no redispatch or load shedding 
 and $P_m$ and $P_n$ do not change when the line outages.
 The term $B_{{m}n}^{(i)}{(B^{(i)}_{nn}})^{-1}P_{n}$ describes how the injected powers $P_n$ 
 redistribute to equivalent injections at the border buses after line $i$ outages,
 and is usually close to the equivalent injections $B_{mn}B_{nn}^{-1}P_{n}$ before the outage.$^{3}$
 Now we consider the effect of the line outage on the power $P_m^{\rm into}$ entering the area R along the tie lines.
 There are two cases.  In the first case, there is no alternative path for power to flow around the area 
 (that is, the area is a cutset area \cite{DobsonHICSS10,DobsonPESGM10} in that removing the area disconnects the network),
 and the power entering the area along the tie lines does not change so that $P_m^{{\rm into}(i)}=P_m^{\rm into}$.
 In the second case, there is an alternative path for the power to flow around the area,
 and  $P_m^{{\rm into}(i)}$ will be different than $P_m^{\rm into}$.
 However, in the practical cases considered in this paper, the alternative paths have fairly high impedance so that the difference 
 between $P_m^{{\rm into}(i)}$ and $P_m^{\rm into}$ is small.
 The conclusion is that in this paper, $P^{(i)}_{\rm area}\approx P_{\rm area}$.
  
 Gathering these relationships and approximations together we obtain
 \begin{align}
 \theta^{(i)}_{\rm area}\approx
 \theta^{[i]}_{\rm area}
 =\frac{P^{(i)}_{\rm area}}{b^{(i)}_{\rm area}}\approx 
 \frac{P_{\rm area}}{b^{(i)}_{\rm area}}
 \label{pareaiex2}
 \end{align}
Also a numerical example of approximation (\ref{pareaiex2}) is given at the end of Section~IV.
Thus  $\theta^{(i)}_{\rm area}$ and $b^{(i)}_{\rm area}$ are approximately inversely related.

\section{Simple examples}
To better understand the relationship between the susceptance of the area and the area voltage angle, we first consider a very simple case of 3 parallel lines connecting bus $a$ to bus $b$ with respective susceptances $b_1$, $b_2$, and $b_3$. Power $P_a$ is generated at bus $a$ and consumed at bus $b$.  In this simple case, the area susceptance $b_{\rm area}=b_1+b_2+b_3$ is the sum of the line susceptances and the area angle 
$\theta_{\rm area}=\theta_a-\theta_b$ is the angle difference between the voltages at bus $a$ and $b$, and the equivalent power  
through the area $P_{\rm area}=P_a$.
In the base case,
\begin{align}
 \theta_{\rm area} =\frac{P_{\rm area}}{b_{\rm area}}=\frac{P_a}{b_1+b_2+b_3}
  \end{align}
If line 1 outages, the power flowing through the area $P_{\rm area}=P_a$ remains constant, the area susceptance 
decreases to  $b_{\rm area}^{(1)}=b_2+b_3$, and the area angle increases to 
 \begin{align}
 \theta_{\rm area}^{(1)}
 = \theta_{\rm area}^{[1]}
 =\theta_a^{(1)}-\theta_b^{(1)}
 =\frac{P_{\rm area}^{(1)}}{b_{\rm area}^{(1)}}
 =\frac{P_a}{b_{\rm area}^{(1)}}
 =\frac{P_a}{b_2+b_3}
 \end{align}
 The voltage angle increase reflects the decreased susceptance in the network and the increased area stress.
 We also have  $\theta_{\rm area}^{(2)}=
 P_a/b_{\rm area}^{(2)}$ and $\theta_{\rm area}^{(3)}=
 P_a/b_{\rm area}^{(3)}$, and it can be seen that outaging the line with the largest susceptance gives the 
 largest increase in area angle.
 
\begin{figure}[h]
\begin{center}
\includegraphics[width=2.8in]{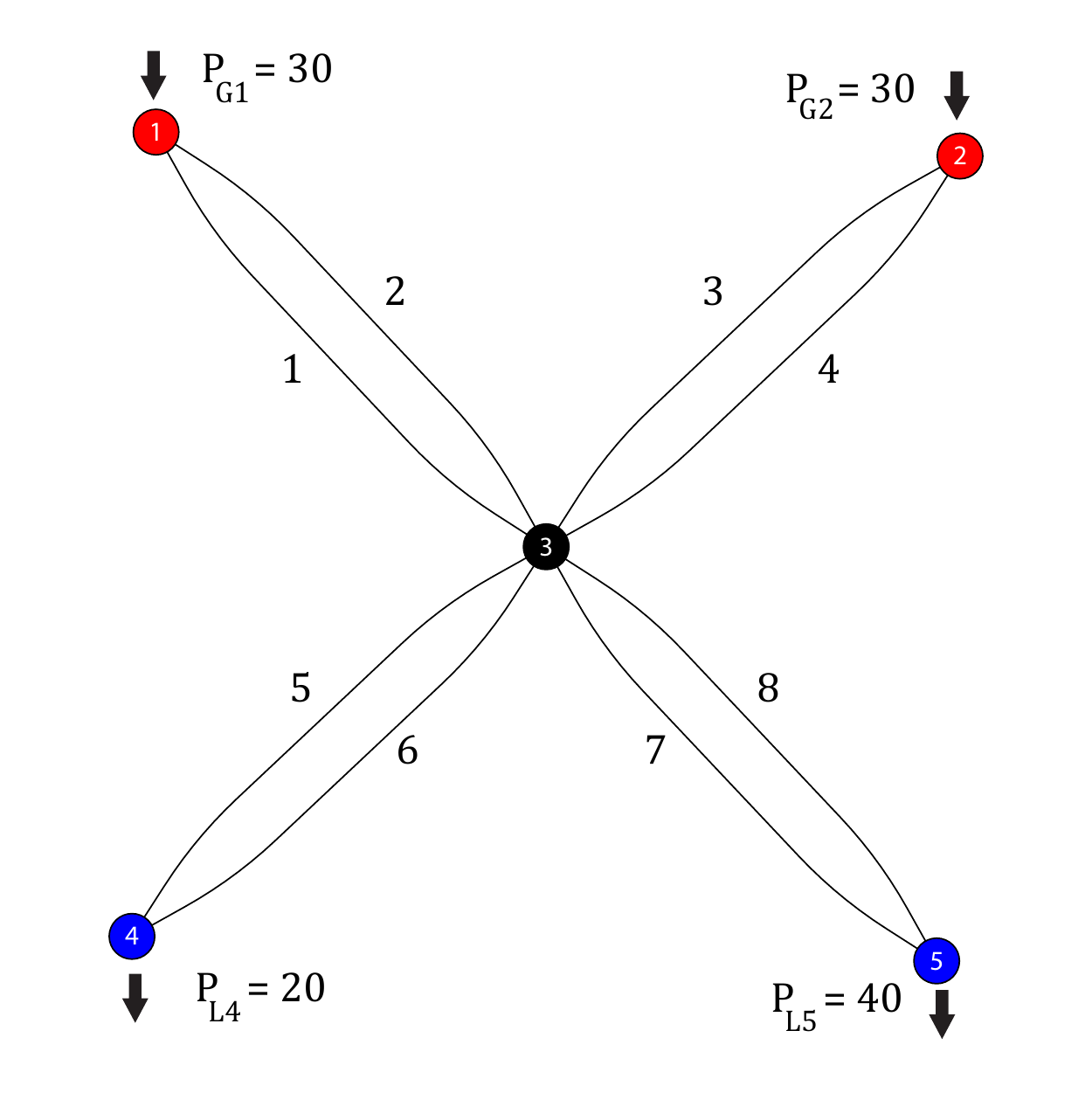}
\caption{5 bus example network with north border buses 1 and 2  in red and south border buses 4 and 5 in blue.}
\label{pic1Change-simpleEx}
\end{center}
\end{figure}

To observe the same  effects in an example in which multiple voltage angles are combined to form the area angle, consider  the simple symmetric 
network shown in  Figure \ref{pic1Change-simpleEx}.
 Buses 1 and 2 are north border buses and buses 4 and 5 are south border buses.  
 The susceptance of  each of the four  lines connected to the north border is 30 pu, the susceptance of each of the two lines between bus  3 and bus  4 is 10 pu, and the susceptance of each of the two lines between bus 3 and bus 5 is 40 pu. The power generation at the north border and the loads at the south border are shown in per unit in Figure \ref{pic1Change-simpleEx}.
 The larger susceptance lines 7 and 8 have a larger power flow of 40 pu.
 
 We are interested in the voltage angle across the area from the north border to the south border, which is the following weighted combination of the 
 border voltage angles:
 \begin{align}
\theta_{\rm area5bus}
=0.5 \, \theta_1+0.5 \, \theta_2-0.33\,  \theta_4-0.67 \,\theta_5
\label{theta1-simpleEx}
\end{align}

\begin{figure}[h]
\begin{center}
\includegraphics[width=\columnwidth]{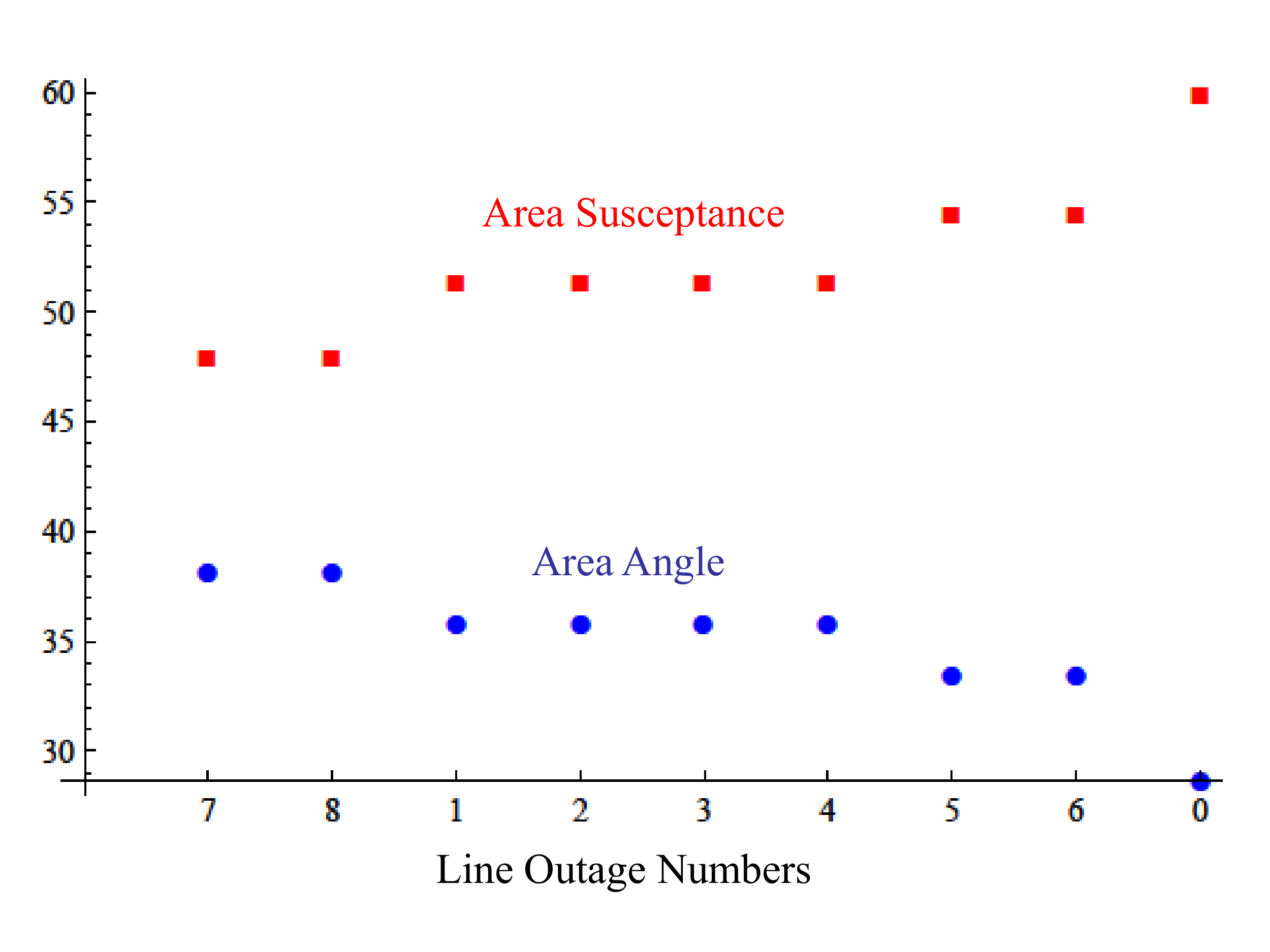}
\caption{Area  angle $\theta_{\rm area5bus}^{(i)}$ in degrees and area susceptance $b_{\rm area5bus}^{(i)}$  in pu for each line outage of 5 bus system. Base case is indicated as line 0.}
\label{pl1Change-simpleEx}
\end{center}
\end{figure}

We take out each line in the system in turn and calculate the area susceptance $b_{\rm area5bus}^{(i)}$ and the monitored area angle $\theta_{\rm area5bus}^{(i)}$ in each case.
The results in Figure \ref{pl1Change-simpleEx} show that the area voltage angle responds to and changes inversely with the area susceptance. 
Moreover, the changes are largest for most severe  line outages. 
For example,  lines 7 and 8  have the largest susceptances and power flows, and when either line 7 or line 8 outages, the area angle increases the most and the susceptance decreases the most.
Lines 5 and 6  have the smallest susceptances and  power flows, and when either line 5 or line 6 outages, the area angle increases the least and the susceptance decreases the least.

\begin{figure}[h]
\begin{center}
\includegraphics[width=2in]{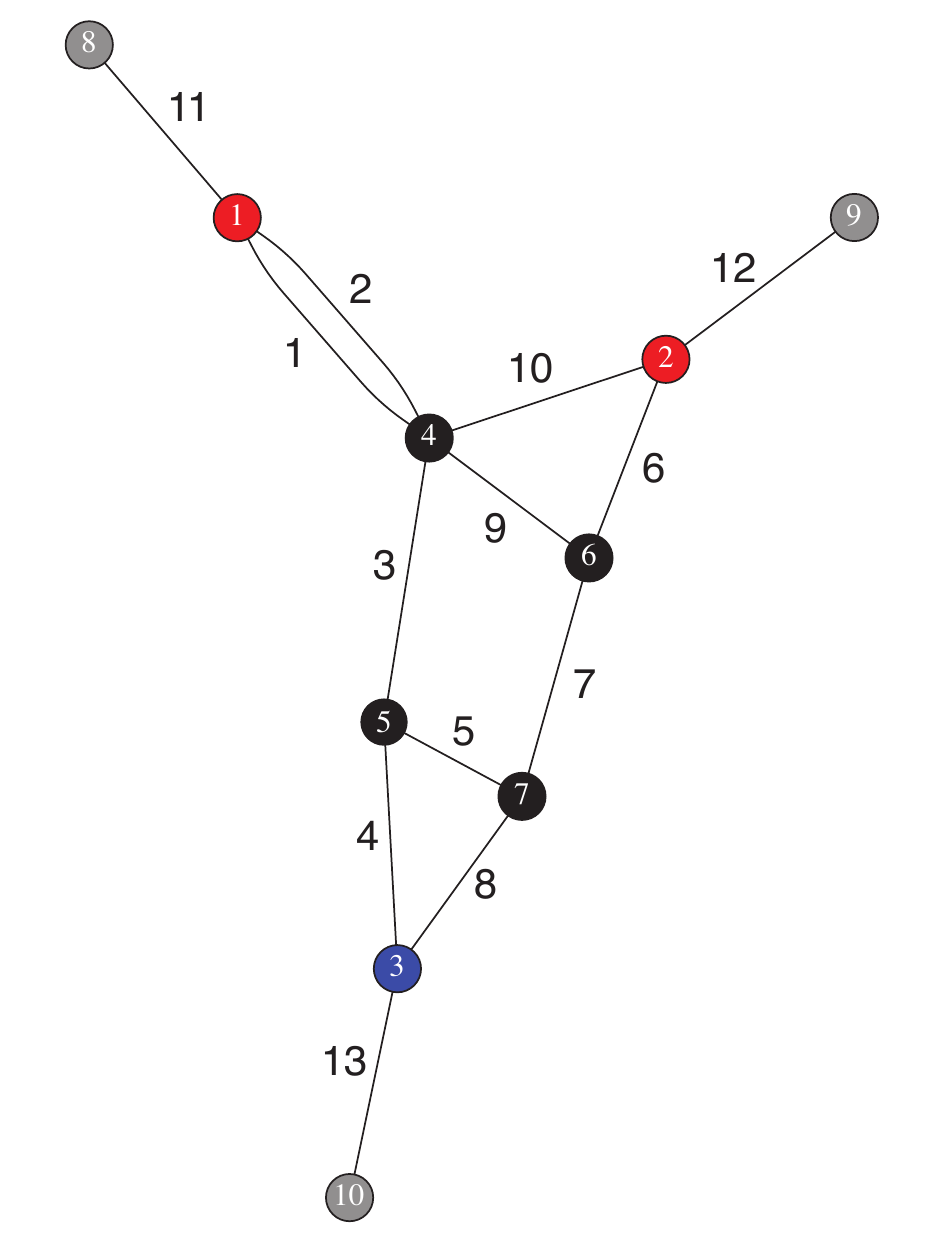}
\caption{9 bus example network with north border buses 1 and 2  in red and the south border bus 3 in blue. The buses inside the area are black.}
\label{pic2Change-simpleEx.jpg}
\end{center}
\end{figure}

A 9 bus example of an asymmetric network with lines of equal susceptance is shown in Figure \ref{pic2Change-simpleEx.jpg}.
 Buses 1 and 2 are north border buses and bus 3 is the south border bus. Buses  8 and 9 have generators each providing 8 pu and bus 10 has load of 16 pu, so the total power into the area at the north border is 16 pu.
 The north to south area angle is  
 \begin{align}
\theta_{\rm area9bus}
=0.44\,  \theta_1+0.56\,  \theta_2-  \theta_3
\label{theta2-simpleEx}
\end{align}

\begin{figure}[h]
\begin{center}
\includegraphics[width=\columnwidth]{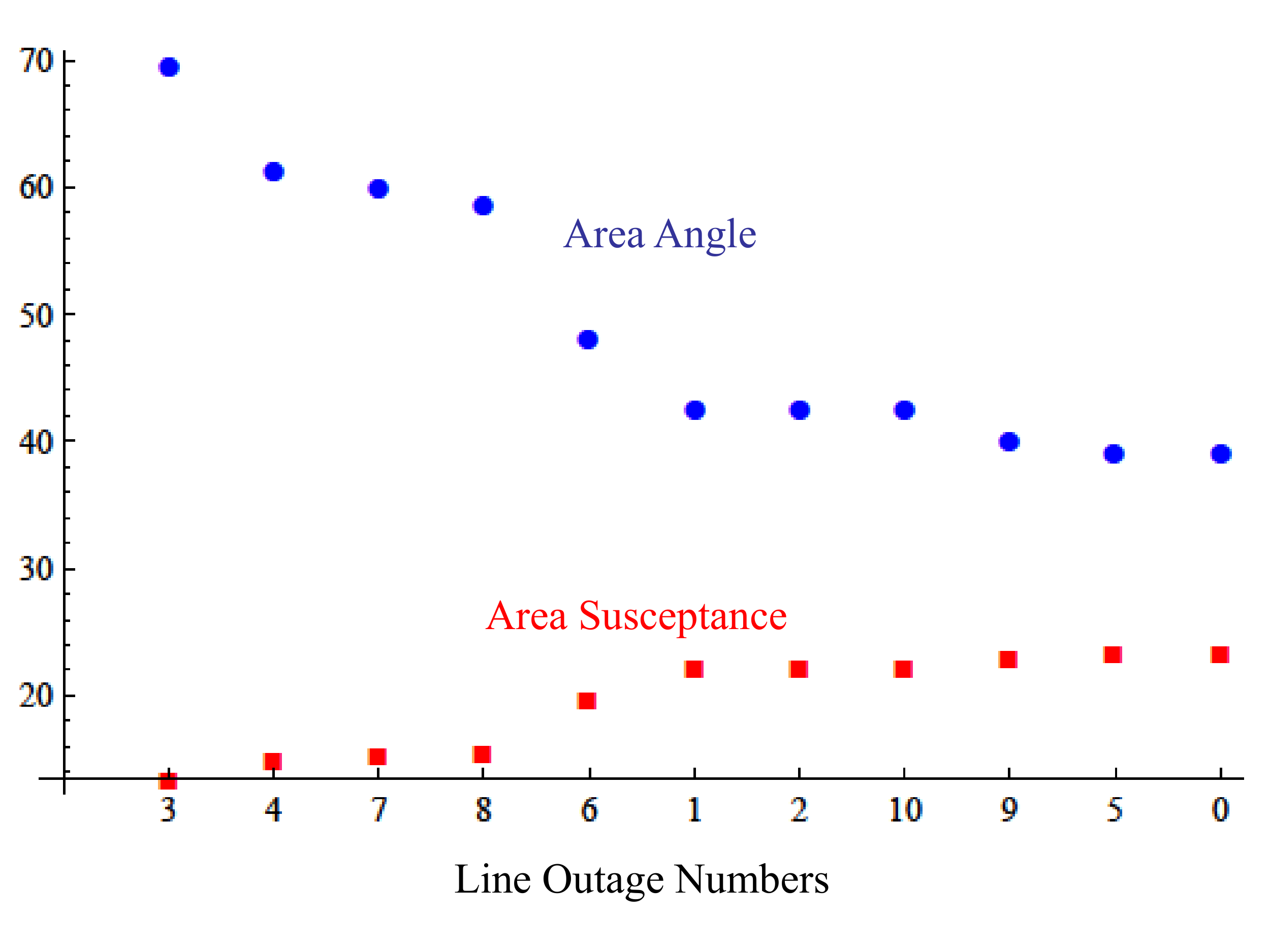}
\caption{Area angle $\theta_{\rm area9bus}^{(i)}$ in degrees and  area susceptance $b_{\rm area9bus}^{(i)}$ in pu for each line outage for 9 bus system. Base case is indicated as outage 0.}
\label{pl2Change-simpleEx}
\end{center}
\end{figure}

The results in Figure \ref{pl2Change-simpleEx} show that the area  angle $\theta_{\rm area9bus}^{(i)}$ responds to and changes inversely with the area susceptance $b_{\rm area9bus}^{(i)}$. 
In this example, although all the lines have the same susceptance, they participate differently in transferring power north to south through the area.
Therefore their outages have different severities and different impacts on the area susceptance and area angle.
For example, 
after the line outages 3, 4, 7, 8, which have the largest power flow since they are in the main path of transferring power from  north to south, we get the largest decrease in the susceptance and also the largest increase in the area angle, which correctly indicates that these are severe outages. In contrast, after the line outages 5 and 9, which have the smallest power flow since they are not in the main path of transferring power from north to south but instead run from east to west, we get the smallest change in area susceptance and area angle,  which correctly indicates that these are less severe outages.

\section{Results for angles across areas of WECC}
\label{results}

We illustrate the use of area angles to monitor single, non-islanding line outages inside two areas of the  WECC system.

\begin{figure}[]
   \begin{center}
   \includegraphics[width=\columnwidth]{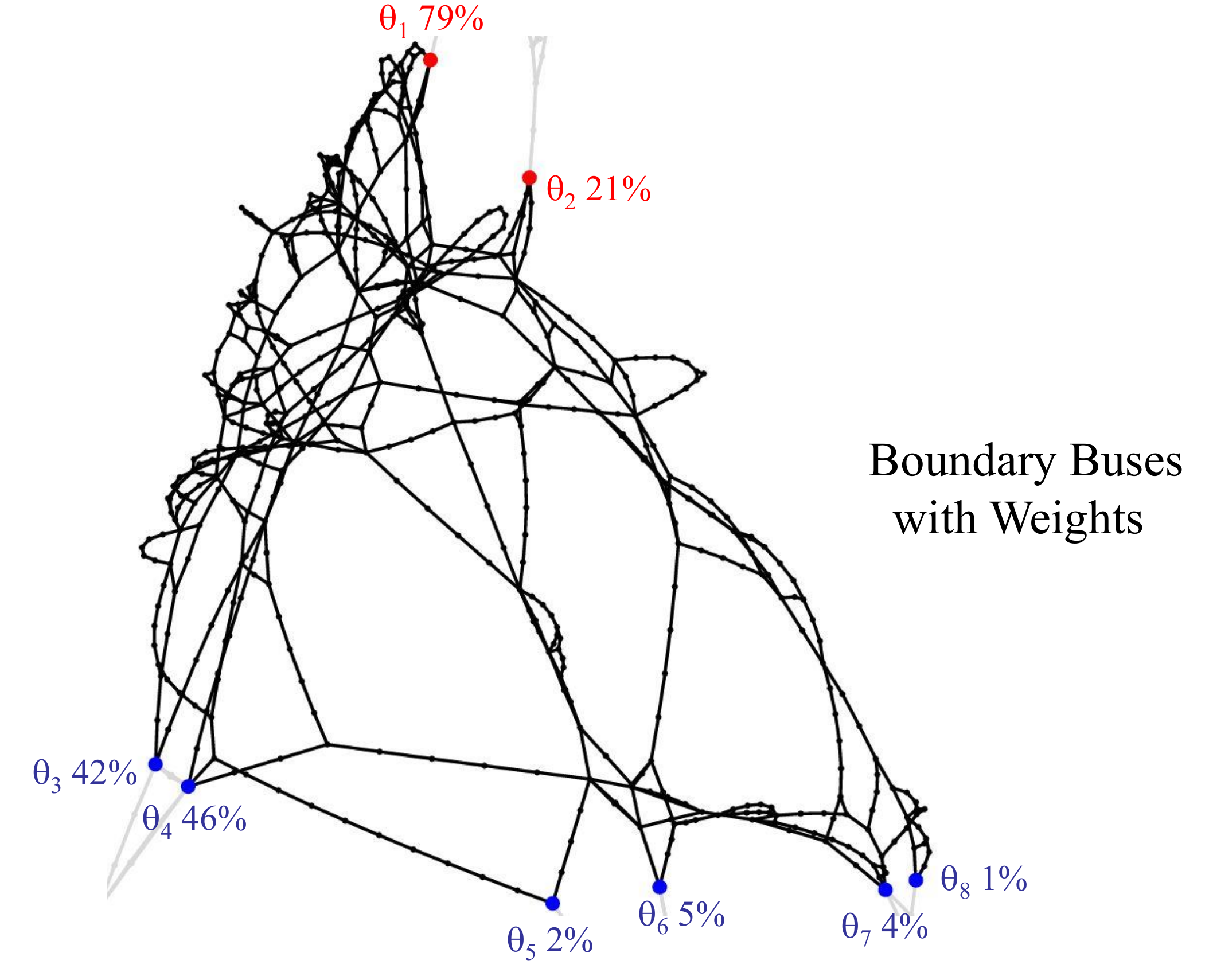}
   \caption{Area 1 of WECC system with area lines in black, north border buses in red, and south border buses in blue. Layout detail is not  geographic.}   \label{pic1Change-WeccIDGeneral}
   \end{center}
   \end{figure}

The first area, for which the network, border buses, and weights are shown in Figure~\ref{pic1Change-WeccIDGeneral}, covers roughly Washington, Oregon, Idaho, Montana, and Wyoming and contains over 700 lines. The north border is near the Canadian border and the 
south border is near the Oregon-California border and its extension eastwards.
The  area angle is the following weighted combination of the border bus angles:
    \begin{align}
\theta_{\rm area1} =\, & 0.79\,  \theta_1 + 0.21\,  \theta_2 - 0.42 \,\theta_3 - 0.46 \,\theta_4\notag\\&
   - 0.02 \,\theta_5 - 0.05 \,\theta_6- 0.04 \,\theta_7 - 0.01 \,\theta_8 \notag
  \end{align}

\begin{figure}[h]
  \begin{center}
  \includegraphics[width=\columnwidth]{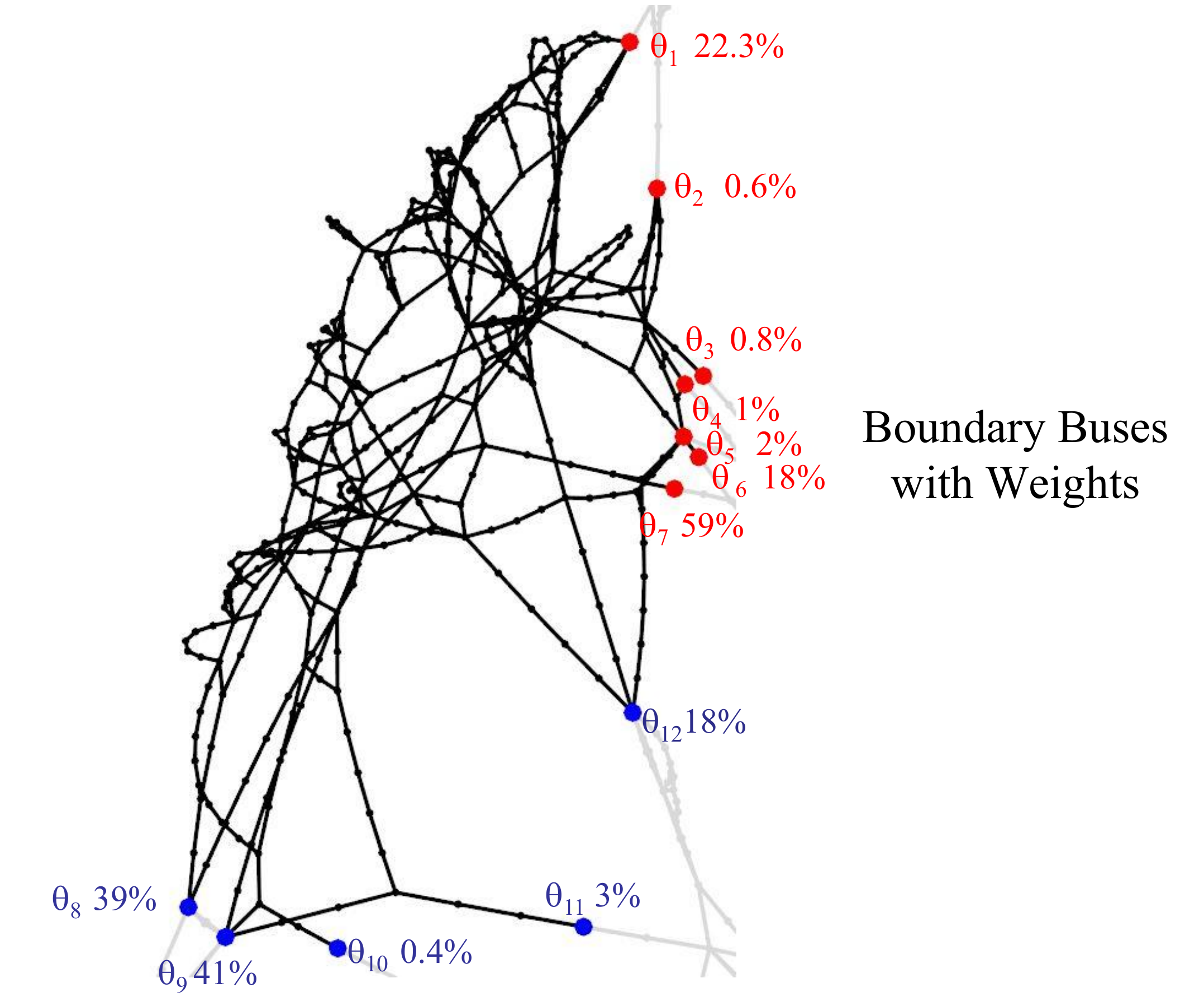}
  \caption{Area 2 of WECC system with area lines in black, north border buses in red and south border buses in blue. Layout detail is not  geographic.}
  \label{pic4Change-WeccIDGeneral}
  \end{center}
  \end{figure}

The second and smaller area shown in  Figure~\ref{pic4Change-WeccIDGeneral} covers roughly Washington and Oregon. The northern (and western) border is near the borders of Canada-Washington,  Washington-Montana and Oregon-Idaho, and the south border is near the Oregon-California border.
       The  area angle is 
    \begin{align}
\theta_{\rm area2} &=\,  0.223\,  \theta_1 + 0.006\,  \theta_2\notag\\& + 0.008 \,\theta_3 + 0.01 \,\theta_4
   + 0.02 \,\theta_5 + 0.18 \,\theta_6+ 0.59 \,\theta_7\notag\\& - 0.39 \,\theta_8 
      - 0.41 \,\theta_9 - 0.004 \,\theta_{10}- 0.03 \,\theta_{11} - 0.18 \,\theta_{12}\notag
  \end{align}
In practice the measurements with very small weights could be omitted.

 For both areas, we are interested in monitoring the north-south area stress with the area angle
when there are single non-islanding line outages, and relating changes in the area angle to the area 
susceptance and the outage severity.
We take out each line in the system in turn and calculate  the monitored area angle $\theta_{\rm area}^{(i)}$ and the area susceptance $b_{\rm area}^{(i)}$ in each case.

\looseness=-1
To quantify the severity of each outage, we compute the maximum power that can enter the area after the outage of each line; for more detail see
\cite{DarvishiNAPS13}. 
The real power through the area is increased by increasing the power entering at each border bus proportionally.
(Generally power enters the area at the northern border buses and leaves the area from the south border buses.)
The maximum power entering the area through the north border occurs when the first line limit inside the area is encountered.
The idea is that the more severe line outages will more strictly limit the maximum power that 
can be transferred north to south through the area. This definition of outage severity can be related to the economic effect 
of limiting the north-south transfer.

\begin{figure}[h]
  \begin{center}
  \includegraphics[width=\columnwidth]{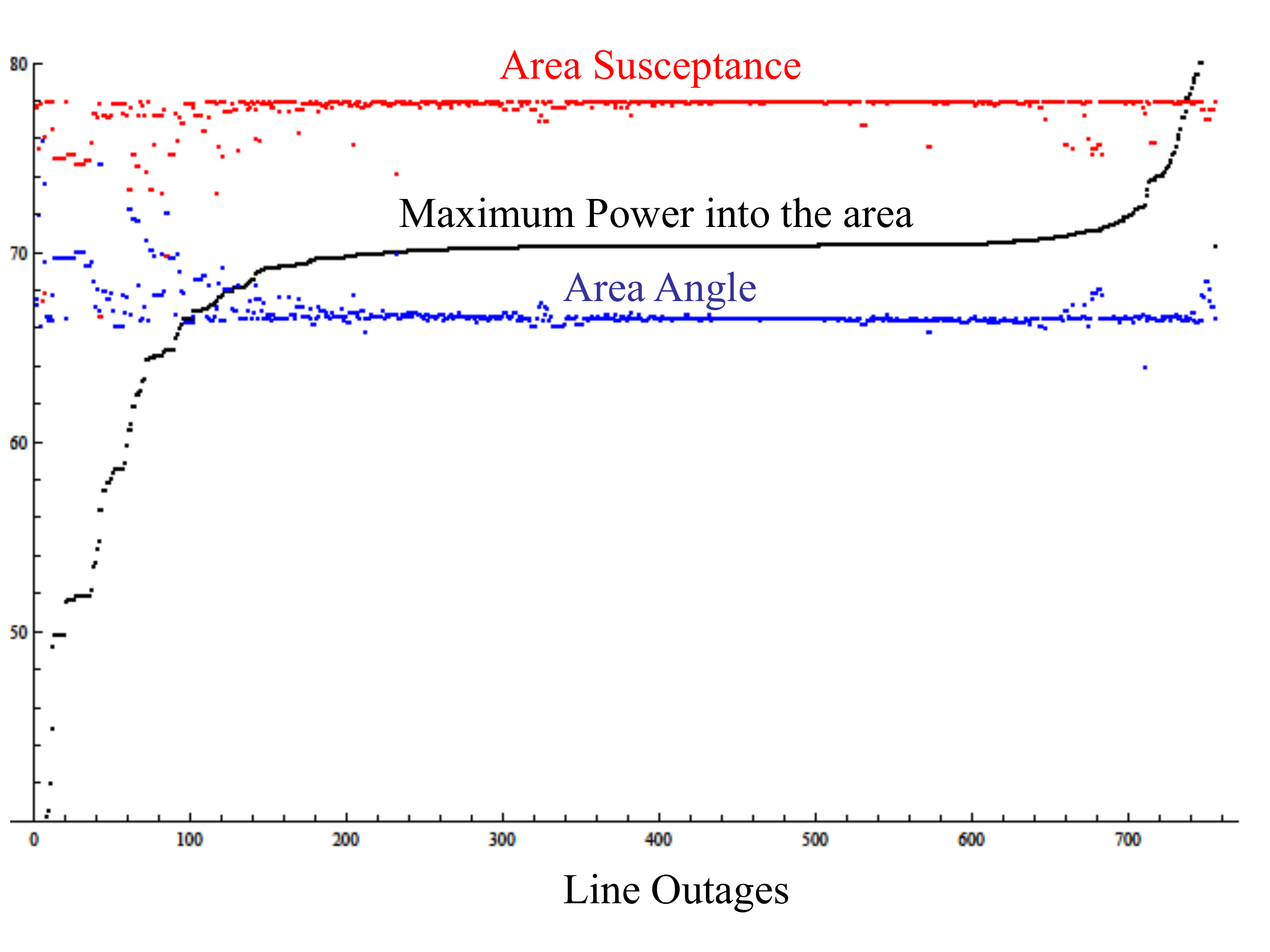}
  \vspace{-25pt}
   \caption{Area angle $\theta_{\rm area1}^{(i)}$ in degrees, area susceptance $b_{\rm area1}^{(i)}$ , and maximum power into the area in pu for each line outage in WECC area 1. Base case (the point at extreme right) is $\theta_{\rm area1}=66.5^{\rm o}$, $b_{\rm area1}=39.0\,$pu, max power  = 46.9.
   For clarity, graph shows $b_{\rm area}$ multiplied by 2, and max power  multiplied by 1.5.}
  \label{pl1Change-WeccIDGeneral}
  \end{center}
  \begin{center}
  \includegraphics[width=\columnwidth]{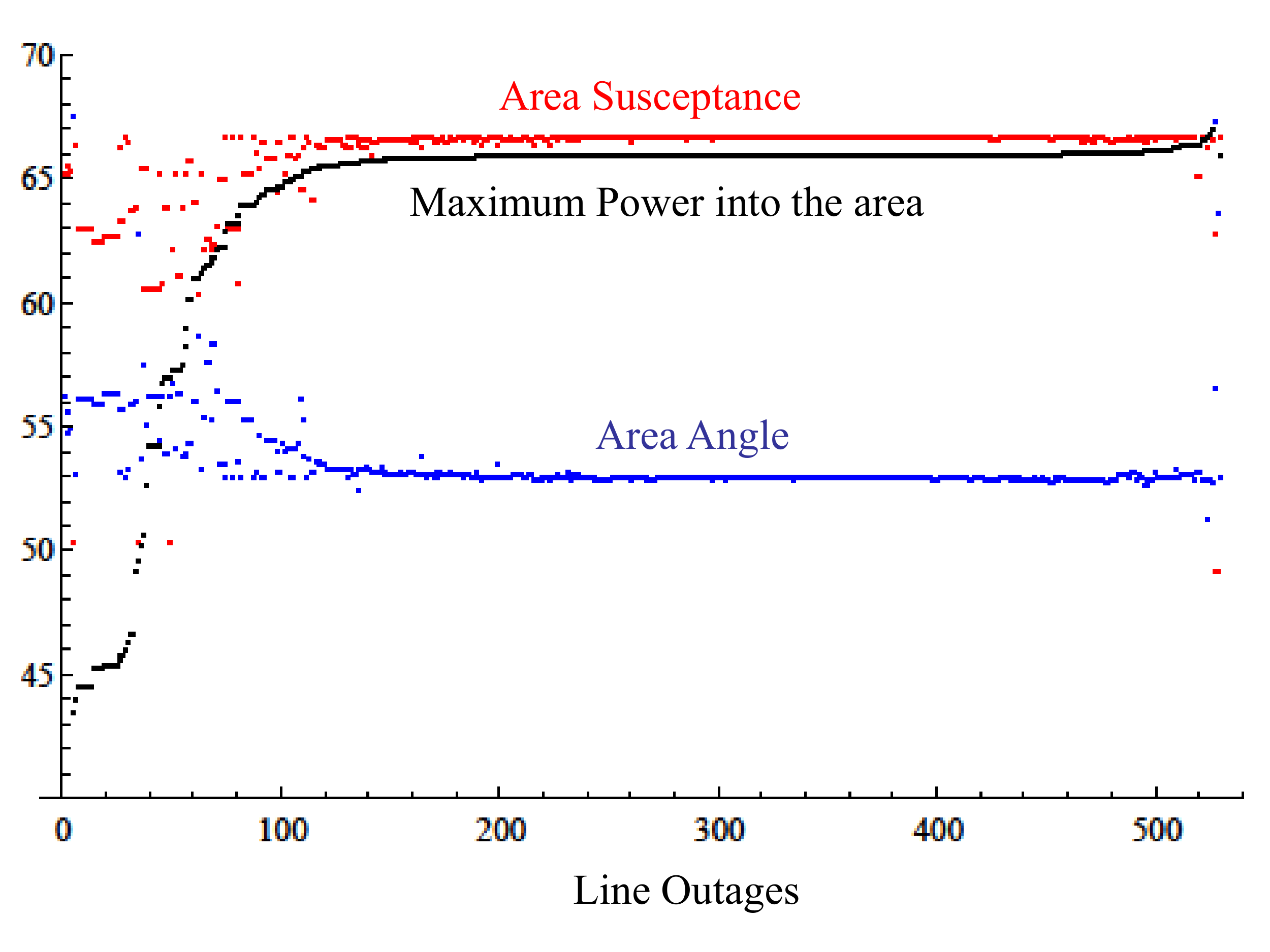}
  \vspace{-25pt}
   \caption{Area angle $\theta_{\rm area2}^{(i)}$ in degrees, area susceptance $b_{\rm area2}^{(i)}$, and max power into the area in pu for each line outage in WECC area 2. Base case (the point at extreme right) is $\theta_{\rm area2}=52.9^{\rm o}$, $b_{\rm area2}=66.7\,$pu, max power  = 66.0\,pu.}  
  \label{pl4Change-WeccIDGeneral}
  \end{center}
  \end{figure}

  The area angle and the area susceptance for each line outage are shown in Figure \ref{pl1Change-WeccIDGeneral} for area 1 and in Figure \ref{pl4Change-WeccIDGeneral} for area 2. 
    The similar patterns of changes in the area angles and area susceptances confirm that the inverse relationship between area angle and area susceptance usually applies.

  \looseness =-1
 Figures~\ref{pl1Change-WeccIDGeneral} and  \ref{pl4Change-WeccIDGeneral} also show the outage severity computed as the maximum power into the area.
  Note that the line outages are sorted according to increasing  maximum power into the area (decreasing  severity). The most severe line outages are on the left hand sides of Figures~\ref{pl1Change-WeccIDGeneral} and  \ref{pl4Change-WeccIDGeneral}, and it 
  can be seen that the area angle usually increases substantially for  most of the severe line outages.  Moreover, in the middle portion of the figures with small changes in severity from the base case (the flat portion of the maximum
  power into the area), the change in area angle from the base case is usually also small.  
  This suggests, for our chosen quantification of outage severity, that large increases in area angle 
  usually indicate the severe line outages. 
  In our experience, this good  result relies on our use of  realistic line limits.
   This tracking of the severity of the outages with the area angle is imperfect, but 
    this is to be expected when trying to monitor over 700 lines in WECC area 1 and 500 lines in WECC area 2 with one scalar area angle as a single bulk area index. (Also note that we are only using a dozen or fewer synchrophasor measurements 
    to compute the area angle.)
   There are several reasons for the exceptional line outages in which the changes in area angle do not track the outage severity.
    Large generation or load inside the area can influence the maximum power entering the area under single line outage conditions,
    and the discrepancy can arise from inaccurate assessment of the outage severity with the maximum power entering the area.
    The line limits that determine the maximum power entering the area and the outage severity
   may not follow the susceptance of the lines and so the susceptance of the area and hence in these cases the area angle cannot track the outage severity. These effects are also the likely cause of the outages at the right of Figure  \ref{pl4Change-WeccIDGeneral} having a maximum power into the area larger than the base case.    

  \looseness=-1
To numerically check the assertion that  $\theta_{\rm area}^{(i)}$ and $\theta_{\rm area}^{[i]}$  are close,  we compute the ratio $\theta_{\rm area}^{(i)}/\theta_{\rm area}^{[i]}$ for each line outage. 
For WECC area 1,
 $\theta_{\rm area1}^{(i)}/\theta_{\rm area1}^{[i]}$ has mean  0.9999, standard deviation 0.002501, and it ranges from 0.9846 to 1.014.
For WECC area 2,
 $\theta_{\rm area2}^{(i)}/\theta_{\rm area2}^{[i]}$ has mean  0.9993, standard deviation 0.006082, and it ranges from 0.9236 to 1.056.

\section{Conclusion}
\label{conclusion}

 It is useful to monitor area angle by combining together synchrophasor measurements at the borders of a suitably chosen area.
The area angle and the area susceptance change when single, non-islanding line outages occur and we show that area angle and susceptance tend to change inversely
using both simple examples and two examples of areas  with hundreds of lines in a real power system.
This approximate relation between area angle and area susceptance gives intuition about how the area angle works
to detect line outages in the area.

The area angle results in a real power system also show that the amount of change in the area angle usually indicates the severity of the line outage
(the exceptions generally relate to outages of lines that are connected to generation or load inside the area).
This suggests that a threshold for changes in the area angle to distinguish severe single line outages could be set.

\newpage
\section*{Acknowledgments}
\label{ack}
We gratefully acknowledge support in part from 
DOE project ``The Future Grid to Enable Sustainable Energy Systems," an initiative of PSERC, 
and
NSF grant CPS-1135825, and the  Electric Power Research Center at Iowa State University.
We gratefully acknowledge access to the WECC data that enabled this research.
The analysis and conclusions are strictly those of the authors and not of WECC.

\newpage

\end{document}